# Mid-infrared ultra-short mode-locked fiber laser utilizing topological insulator $Bi_2Te_3$ nano-sheets as the saturable absorber

Ke Yin[1], Tian Jiang[1,2,3], Hao Yu[1], Xin Zheng[1], Xiangai Cheng[1,2], Jing Hou[1,4]

[1]College of Optoelectronic Science and Engineering, National University of Defense Technology, Changsha 410073, P.R.China;
[2]State key laboratory of High Performance Computing, National University of Defense Technology, Changsha 410073, P.R.China;
[3] jiangtian198611@163.com
[4]houjing25@sina.com

**Abstract:** The newly-emergent two-dimensional topological insulators (TIs) have shown their unique electronic and optical properties, such as good thermal management, high nonlinear refraction index and ultrafast relaxation time. Their narrow energy band gaps predict their optical absorption ability further into the mid-infrared region and their possibility to be very broadband light modulators ranging from the visible to the mid-infrared region. In this paper, a mid-infrared mode-locked fluoride fiber laser with TI Bi2Te3 nano-sheets as the saturable absorber is presented. Continuous wave lasing, Q-switched and continuous-wave mode-locking (CW-ML) operations of the laser are observed sequentially by increasing the pump power. The observed CW-ML pulse train has a pulse repetition rate of 10.4 MHz, a pulse width of ~6 ps, and a center wavelength of 2830 nm. The maximum achievable pulse energy is 8.6 nJ with average power up to 90 mW. This work forcefully demonstrates the promising applications of two-dimensional TIs for ultra-short laser operation and nonlinear optics in the mid-infrared region.

## 1. Introduction

Current researches in new optical functional devices and saturable absorber (SAs) primarily focus on layered structure materials [1, 2]. The advancement of graphene in SA motivates the exploration of other layered structure materials [3, 4], such as topological insulators (TIs) [5-7] and transition metal dichalcogenides [8]. Although the zero band gap of graphene shows that it may be fabricated as an ultra-broad optoelectronics devices which has ultra-wide photon response, but its intrinsic defections or limitations such as the weak light absorption, low damage threshold and low saturing intensity make it less attractive for specific applications. TIs are a new class of materials which have a bulk band gap with gapless Dirac surface/edge states, protected by their topological symmetry. Originating from the combined effects of spin-orbit interactions and time-reversal symmetry [6], TIs own unique electronic and optical properties such as good thermal management, high nonlinear refraction index and ultrafast relaxation time. Since F. Bernard et al. firstly reported the saturable absorption of TI $Bi_2Te_3$ in 2012 [9], plenty of efforts and trials have demonstrated the utilization of different TI-SAs for its nonlinear light absorption ability in either passively mode-locked (ML) lasers [10-17] or Q-switched (QS) lasers [18-22] in near-infrared region. In fact, the energy band gaps of most TIs are 0.2~0.3 eV [23], which correspond to their broadband nonlinear optical response from the visible to the mid-infrared region (wavelengths shorter than 4.1 μm). Previous works have shown the possibility of TIs' application in mid-infrared pulsed lasers rather than merely be constrained in the near-infrared region.

Laser sources in the mid-infrared region are useful for applications such as laser surgery, spectroscopy, nonlinear optics, environmental monitoring and defense [24-27]. But, ordinary silica fibers show their inability in the mid-infrared region due to silica's strong intrinsic multi-phonon absorption at wavelengths longer than 2.4 μm [28]. People have to look for glass materials which are transparent in the mid-infrared region. These materials include fluorides, chalcogenide glasses, tellurite glasses, etc. As one type of fluoride fibers, rare earth doped $ZrF_4$-$BaF_2$-$LaF_3$-$AlF_3$-$NaF$ (ZBLAN) fibers have been widely investigated in mid-infrared fiber lasers [29-33], showing their high stability, elegant thermal management and comparative good environmental durability. There are many reports on mid-infrared fiber lasers incorporating with either $Er^{3+}$ or $Ho^{3+}$ doped ZBLAN fibers with output wavelengths ranging from 2.7 to 3 μm. In particular, QS or ML laser pulses with high peak power are often more useful in many specific applications than CW fiber laser considering their ability to afford much high peak power. Historically, mid-infrared pulsed fiber lasers have been realized by incorporating either active modulators (acoustic-optic modulators, electro-optic modulators) or passive saturable absorbers (semiconductor saturable absorber mirror [33], graphene [34, 35], $Fe^{2+}$: ZnSe crystal [35], InAs crystal [36]) into the laser cavities. Comparatively, passive saturable absorbers afford much greater compactness, better stability, and easier manipulation and maintenance than active method. As aforementioned, it is possible to apply TI-SAs into the mid-infrared pulsed lasers, but there are not any reports yet.

In this paper, we present a mid-infrared ultra-short mode-locked fiber laser at 2830 nm by utilizing a TI saturable absorber mirror (TI-SAM) as the mode-locker. The TI-SAM is fabricated by drop casting TI $Bi_2Te_3$ nano-sheets onto the surface of a gold-coated mirror. The laser has a linear cavity where the TI-SAM serves as the mode-locker and the high reflection mirror. The fluoride fiber laser has a pump threshold of ~120 mW at 1150 nm and a fitted lasing slope efficiency of ~8.9%. Both Q-switched mode-locked (QS-ML) and continuous-wave mode-locked (CW-ML) pulses were realized in the experiments. The obtained CW-ML pulses have a repetition rate of 10.4 MHz, a pulse width of ~6 ps, and a center wavelength of 2830 nm. To our best knowledge, this is the first time to report TIs' application as an optical SA in the mid-infrared fiber lasers.

## 2. Experimental preparation and laser setup

### 2.1. Preparation of TI $Bi_2Te_3$ nano-sheets

Many techniques have been applied to fabricate TI $Bi_2Te_3$ based SA, including liquid or mechanical deposition onto tapered or side-polished fibers, fiber end facet, and through drop cast onto a quartz plate or mirror, etc. In a simple synthesis, a moderate ethylene glycol was used to dissolve a stoichiometric ratio of bismuth chloride ($BiCl_3$) and sodium selenide ($Na_2TeO_3$) together with vigorous stirring. The mixture was then transferred into an autoclave to obtain gray $Bi_2Te_3$ powders, washed with distilled water and ethanol, and finally dried at 330 K in vacuum. The as-grown and washed powders was then were dispersed in an ethanol solution. The initial dispersions were treated for 10 min by a powerful ultrasonic cleaner.

After sonication, the dispersions were allowed to settle for several hours. To extract uniform TI $Bi_2Te_3$ nano-sheets, the upper supernatant was collected with an injector. Later, the dispersion was dropped cast onto a gold-coated mirror which served as the TI-SAM, which was afterwards put into a drying oven for evaporation over 12 h.

### 2.2. Characteristics of TI $Bi_2Te_3$ nano-sheets

After the preparation of TI $Bi_2Te_3$ nano-sheets, many methods are used to characterize these nano-materials. Figure 1 shows the experimental results. The morphology and size of the as-prepared TI $Bi_2Te_3$ nano-sheets were characterized with a scanning electron microscope (SEM) with different scales as shown in Fig. 1(a) and 1(b). From Fig. 1(b), it is obvious that the as-prepared BT nano-sheets have uniform hexagonal morphologies with dimensions between 400 and 600 nm. The thickness and dimensions of the TI $Bi_2Te_3$ nano-sheets were also characterized by using an atomic force microscope (AFM), and the scanned height image of a typical hexagonal flake is presented in the inset of Fig. 1(c). The curve plotted in Fig. 1(c) depicts the measured height data from point m to n, showing that the single flake had a thickness from 30 to 40 nm. These results are in good agreement with previous works about TI $Bi_2Te_3$ nano-sheets.

In order to characterize the wideband optical response of TI materials at the mid-infrared region, a mid-infrared supercontinuum source spanning from 1900 nm to 3500 nm was used to measure the linear transmission of prepared BT nano-sheets. At this time, the TI dispersion was dropped cast onto a thin $CaF_2$ window plate. The calculated mid-infrared transmission of TI materials is plotted in Fig. 1(d), indicating a broadband linear absorption response. Figure 1(e) shows the photograph of the as-prepared TI-SAM, where the gold-coated mirror has a diameter of 25.4 mm. Figure 1(f) gives a 1000 manifolds photo view of the TI deposition area on the surface of the mirror with an optical microscope.

### 2.3. Laser setup

The experimental setup is shown in Fig. 2. The laser had a linear cavity configuration with a 1150 nm fiber laser served as the pump, a piece of double-cladding $Ho^{3+}Pr^{3+}$ co-doped

ZBLAN fiber (HoPr-ZBLAN) to provide the gain, a gold mirror as the high reflectivity mirror, and the perpendicular cleaved fiber end with 4% reflectivity as the output coupler. The as-prepared TI-SAM was used to be the mode-locker and the high reflectivity mirror mounted onto a five-axis positioner. The HoPr-ZBLAN fiber had a length of ~8.5 m. The core of the fiber had a diameter of 10 μm and numerical aperture (NA) of 0.20, and the inner-cladding had a shape of octagon with diameter of 125 μm and NA of 0.46. The gain fiber had a single mode cutoff wavelength of 2.6 μm. The dopant concentrations of $Ho^{3+}$ and $Pr^{3+}$ in the fiber core were 30000 ppm and 2500 ppm molar, respectively. A stable home-made 1150 nm $Yb^{3+}$-doped fiber laser was adopted as the pump light. A customized dichroic mirror (DM) was placed at an angle of 45º, with a designed transmittance of >95% at 1150 nm and a designed reflectance of ~99.5% at 3 μm. The collimated pump light was coupled into the perpendicularly cleaved fiber end through a silica lens with a focal length of 50 mm. The other end of the HoPr-ZBLAN fiber was cleaved at an angle of 9º, and coupled onto the Ti-SAM with two aspheric lenses to provide feedback. The output laser was collimated by two $CaF_2$ lenses to the monitoring system. The two $CaF_2$ lenses were anti-reflection coated at 3 μm.

The measuring system includes a thermal power meter, a monochromator together with a liquid-nitrogen-cooled InSb detector, an interference autocorrelator. A 20 GHz sampling rate digital oscilloscope with 1.5 GHz bandwidth and a photovoltaic infrared detector were used to monitor the output pulses. The high cut-off frequency of the photovoltaic infrared detector is limited to 250 MHz.

## 3. Results and discussion

The performance of the laser was primarily tested by adjusting the position of gold-coated mirror, but only the region which had no TI materials as the reflected region. This results in the CW operation of the fluoride fiber laser, with a linear fitted power slope efficiency of 10%. Then the reflection position of gold-coated mirror was adjusted to let TI $Bi_2Te_3$ materials could be incorporated into the laser cavity. At this time, the lasing threshold of the laser got a little higher and the fitted signal to pump power slope efficiency decreased to ~8.9%. The measured output power was scaled almost linearly along with the increasing of the 1150 nm pump power. When the pump power was adjusted between 114 and 370 mW, only CW lasing at 2830 nm was measured. While the pump power exceeded 370 mW, obvious temporal instabilities of the output were observed. When pump power was changed from 370 to 680 mW, stable QS-ML pulses were observed on the oscilloscope.

Figure 3 plots three typical measured results of output pulse trains in the QS-ML state at the pump power of 503, 576 and 630 mW. It is evident that QS envelopes appeared on the top of the mode-locked pulses, indicating the state of QS-ML operations. The Q-switched envelopes were very stable and appeared repeatedly. The repetition rate for the QS envelope was 61 and 80 kHz at the pump power of 503 and 576 mW, respectively. By taking a close-up view of the mode-locked pulses, the pulse to pulse interval was measured to be 96 ns which corresponded to a repetition rate of 10.42 MHz and matched with the cavity round-trip time. With the increasing of the pump power, the repetition rate of the Q-switched envelope increased at the same time and trended to disappear. It was found the peak to peak fluctuations of the output pulse trains decreased from 88 % to 19% by strengthening the pump from 503 mW to 630 mW as shown in Fig .3.

Moreover, by strengthening the 1150 nm pump power to higher than 680 mW, CW-ML operation was realized in the experiments. Fig 4(a) plots the measured pulse train when the periodic Q-switched envelope disappeared and CW-ML operation of the fluoride fiber laser was obtained. Fig. 4(b) depicts the single pulse profile, showing a pulse width of 2.1 ns limited by the bandwidth of the photovoltaic infrared detector. It is found that the pulse to pulse intensity fluctuation was further decreased to less than 10% as shown in the inset of Fig. 4(b). Pulse characteristics were kept until the pump power was increased to ~1.1 W. The output spectrum of stable CW-ML pulses is depicted in Fig. 4(c), with a center wavelength of 2830 nm and a 3 dB spectral width of 10 nm. A Michelson interferometer was built to measure the output pulse width. In the experiments, interference signals were detected on the oscilloscope by adjusting the moving arm of the Michelson interferometer over a distance of 3.6 mm, indicating a pulse width of ~ 6 ps. The maximum output power of CW mode-locked pulses was 90 mW, which corresponds to single pulse energy of 8.6 nJ. Therefore, the maximal peak power could be calculated to 1.4 kW.

Usually, further increase of the pump power may lead to much higher output pulse energy operation of the CW-ML state or harmonically mode-locked operation of the ZBLAN fiber laser. However, when we did this trial surface damage of the gold mirror and the ablation of the deposited TI materials were observed in the experiments. The reason for this phenomenon was that the gold film was coated on a polyester substrate which hindered the heat dissipation when the laser is on high power operation. An improvement to overcome this limitation is to replace the gold-coated mirror with other better thermal management mirrors. Although CW-ML pulses were obtained in the experiments, it was believed that the pulse stability could also be improved for future works. There are mainly two aspects for the improvement of pulse stability generated from the fluoride fiber laser which will be our future work, on the one hand, the free-space experimental setup might be replaced with an compact all-fiber laser setup to enhance the environmental durability; on the other hand, necessary polarization control of the laser performance should be considered for long term CW-ML operation.

## 4. Conclusions

In conclusions, we have presented an ultra-short mode-locked fluoride fiber laser at the mid-infrared region incorporating with the newly-emergent two-dimensional TI materials. Uniform TI $Bi_2Te_3$ nano-sheets were drop casted onto the surface of a gold-coated mirror to make the TI-SAM. Both QS-ML pulses and CW-ML pulses were observed. The CW-ML pulses had a pulse repetition rate of 10.4 MHz, a pulse width of ~6 ps, and a center wavelength of 2830 nm. Those results demonstrated the first report of TIs' application as an optical SA for the mode-locked fiber lasers in the mid-infrared region.

**Acknowledgements**

The authors would like to thank Bin Zhang, Linyong Yang and Lei Li at the National University of Defense Technology for their tasteful discussions for the experimental results. The authors also would like the Chujun Zhao and Yu Chen at the Shenzhen University for their supports of the TI materials. This work was supported by by the State Key Program of National Natural Science of China (Grant No. 61235008, 61405254, 61340017 and 61435009) and the Fundamental Researches Foundation of National University of Defense Technology (Grant No. GDJC13-04).

**References**

1. Z. Sun, D. Popa, T. Hasan, F. Torrisi, F. Wang, E. J. R. Kelleher, J. C. Travers, V. Nicolosi, and A. C. Ferrari, "A stable, wideband tunable, near transform-limited, graphene-mode-locked, ultrafast laser," Nano Research 3(9), 653-660 (2010).

2. A. Martinez and Z. Sun, "Nanotube and graphene saturable absorbers for fibre lasers," Nat. Photon. 7(11), 842-845 (2013).

3. J. Wang, "Two-dimensional semiconductors for ultrafast photonic applications," in SPIE OPTO, (Proc. of SPIE, 2015), 9359021.

4. M. Buscema, D. J. Groenendijk, S. I. Blanter, G. A. Steele, H. S. van der Zant, and A. Castellanos-Gomez, "Fast and broadband photoresponse of few-layer black phosphorus field-effect transistors," Nano Lett. 14, 3347-3352 (2014).

5. D. Hsieh, D. Qian, L. Wray, Y. Xia, Y. S. Hor, R. J. Cava, and M. Z. Hasan, "A topological Dirac insulator in a quantum spin Hall phase," Nature 452(7190), 970-974 (2008).

6. H. Zhang, C. Liu, X. Qi, X. Dai, Z. Fang, and S. Zhang, "Topological insulators in Bi2Se3, Bi2Te3 and Sb2Te3 with a single Dirac cone on the surface," Nat. Phys. 5(6), 438-442 (2009).

7. J. E. Moore, "The birth of topological insulators," Nature 464(7286), 194-198 (2010).

8. H. S. S. Ramakrishna Matte, A. Gomathi, A. K. Manna, D. J. Late, R. Datta, S. K. Pati, and C. N. R. Rao, "MoS2 and WS2 Analogues of Graphene," Angewandte Chemie 122(24), 4153-4156 (2010).

9. F. Bernard, H. Zhang, S. Gorza, and P. Emplit, "Towards mode-locked fiber laser using topological insulators," in Nonlinear Photonics, OSA Technical Digest (Optical Society of America, Colorado, 2012), Paper No. NTh1A.5, OSA Technical Digest (online) 2012),

10. C. Zhao, H. Zhang, X. Qi, Y. Chen, Z. Wang, S. Wen, and D. Tang, "Ultra-short pulse generation by a topological insulator based saturable absorber," Appl. Phys. Lett. 101(21), 211106 (2012).

11. Z. Luo, M. Liu, H. Liu, X. Zheng, A. Luo, C. Zhao, H. Zhang, S. Wen, and W. Xu, "2 GHz passively harmonic mode-locked fiber laser by a microfiber-based Topological Insulator saturable absorber," Opt. Lett. 38(24), 5212-5215 (2013).

12. J. Boguslawski, J. Sotor, G. Sobon, J. Tarka, J. Jagiello, W. Macherzynski, L. Lipinska, and K. M. Abramski, "Mode-locked Er-doped fiber laser based on liquid phase exfoliated Sb2Te3 topological insulator," Laser Phys. 24(10), 105111 (2014).

13. C. Chi, J. Lee, J. Koo, and J. H. Lee, "All-normal-dispersion dissipative-soliton fiber laser at 1.06 μm using a bulk-structured Bi2Te3 topological insulator-deposited side-polished fiber," Laser Phys. 24(10), 105106 (2014).

14. M. Jung, J. Lee, J. Koo, J. Park, Y. Song, K. Lee, S. Lee, and J. H. Lee, "A femtosecond pulse fiber laser at 1935 nm using a bulk-structured Bi2Te3 topological insulator," Opt. Express 22(7), 7865-7874 (2014).

15. Y. H. Lin, C. Y. Yang, S. F. Lin, W. H. Tseng, Q. l. Bao, C. I. Wu, and G. R. Lin, "Soliton compression of the erbium-doped fiber laser weakly started mode-locking by nanoscale p-type Bi2Te3 topological insulator particles," Laser Phys. Lett. 11(5), 055107 (2014).

16. J. Sotor, G. Sobon, K. Grodecki, and K. M. Abramski, "Mode-locked erbium-doped fiber laser based on evanescent field interaction with Sb2Te3 topological insulator," Appl. Phys. Lett. 104(25), 251112-1 (2014).

17. K. Yin, B. Zhang, J. Hou, L. Li, X. Zhou, and T. Jiang, "Soliton Mode-Locked Fiber Laser Based on Topological Insulator Bi2Te3 Nanosheets at 2 μm," Photon. Res. 3(3), 72-76 (2015).

18. Z. Luo, Y. Huang, J. Weng, H. Cheng, Z. Lin, B. Xu, Z. Cai, and H. Xu, "1.06 μm Q-switched ytterbium-doped fiber laser using few-layer topological insulator $Bi_2Se_3$ as a saturable absorber," Opt. Express 21(24), 29516-29522 (2013).

19. J. Koo, J. Lee, C. Chi, and J. H. Lee, "Passively Q-switched 1.56 μm all-fiberized laser based on evanescent field interaction with bulk-structured bismuth telluride topological insulator," J. Opt. Soc. Am. B 31(9), 2157-2162 (2014).

20. J. Lee, J. Koo, C. Chi, and J. H. Lee, "All-fiberized, passively Q-switched 1.06 μm laser using a bulk-structured Bi2Te3 topological insulator," J. Opt. 16(8), 085203 (2014).

21. Z. Yu, Y. Song, J. Tian, Z. Dou, H. Guoyu, K. Li, H. Li, and X. Zhang, "High-repetition-rate Q-switched fiber laser with high quality topological insulator Bi2Se3 film," Opt. Express 22(10), 11508-11515 (2014).

22. Z. Luo, C. Liu, Y. Huang, D. Wu, J. Wu, H. Xu, Z. Cai, Z. Lin, L. Sun, and J. Weng, "Topological-Insulator Passively Q-Switched Double-Clad Fiber Laser at 2 μm Wavelength," IEEE J. Sel. Topics Quantum Electron. 20(5), 1-8 (2014).

23. H. Yu, H. Zhang, Y. Wang, C. Zhao, B. Wang, S. Wen, H. Zhang, and J. Wang, "Topological insulator as an optical modulator for pulsed solid-state lasers," Laser Photon. Rev. 7(6), L77-L83 (2013).

24. S. Tokita, M. Murakami, S. Shimizu, M. Hashida, and S. Sakabe, "Liquid-cooled 24 W mid-infrared Er:ZBLAN fiber laser," Opt. Lett. 34(20), 3062-3064 (2009).

25. J. Li, D. D. Hudson, Y. Liu, and S. D. Jackson, "Efficient 2.87 μm fiber laser passively switched using a semiconductor saturable absorber mirror," Opt. Lett. 37(18), 3747-3749 (2012).

26. S. Tokita, M. Murakami, and S. Shimizu, "High Power 3 μm Erbium Fiber Lasers," in Advanced Solid State Lasers, OSA Technical Digest (online) (Optical Society of America, 2014), AM3A.4.

27. S. D. Jackson, "Towards high-power mid-infrared emission from a fibre laser," Nat. Photon. 6(7), 423-431 (2012).

28. J. Lægsgaard and H. Tu, "How long wavelengths can one extract from silica-core fibers?," Opt. Lett. 38(21), 4518-4521 (2013).

29. T. Hu, D. D. Hudson, and S. D. Jackson, "Actively Q-switched 2.9 μm Ho3+Pr3+-doped fluoride fiber laser," Opt. Lett. 37(11), 2145-2147 (2012).

30. J. F. Li, Y. Yang, D. D. Hudson, Y. Liu, and S. D. Jackson, "A tunable Q-switched Ho3+-doped fluoride fiber laser," Laser Phys. Lett. 10(4), 045107 (2013).

31. S. Crawford, D. D. Hudson, and S. Jackson, "3.4 W Ho3+, Pr3+ Co-Doped Fluoride Fibre Laser," in CLEO: 2014, OSA Technical Digest (online) (Optical Society of America, 2014), STu1L.3.

32. J. F. Li, H. Y. Luo, Y. Liu, L. Zhang, and S. D. Jackson, "Modeling and Optimization of Cascaded Erbium and Holmium Doped Fluoride Fiber Lasers," IEEE J. Sel. Topics Quantum Electron. 20(5), 1-14 (2014).

33. J. F. Li, H. Y. Luo, Y. L. He, Y. Liu, L. Zhang, K. M. Zhou, A. G. Rozhin, and S. K. Turistyn, "Semiconductor saturable absorber mirror passively Q-switched 2.97 μm fluoride fiber laser," Laser Phys. Lett. 11(6), 065102 (2014).

34. C. Wei, X. Zhu, F. Wang, Y. Xu, K. Balakrishnan, F. Song, R. A. Norwood, and N. Peyghambarian, "Graphene Q-switched 2.78 μm Er3+-doped fluoride fiber laser," Opt. Lett. 38(17), 3233-3236 (2013).

35. G. Zhu, X. Zhu, K. Balakrishnan, R. A. Norwood, and N. Peyghambarian, "Fe2+: ZnSe and graphene Q-switched singly Ho3+-doped ZBLAN fiber lasers at 3 μm," Opt. Mater. Express 3(9), 1365-1377 (2013).

36. T. Hu, D. D. Hudson, and S. D. Jackson, "Stable, self-starting, passively mode-locked fiber ring laser of the 3 μm class," Opt. Lett. 39(7), 2133-2136 (2014).

**Figures:**

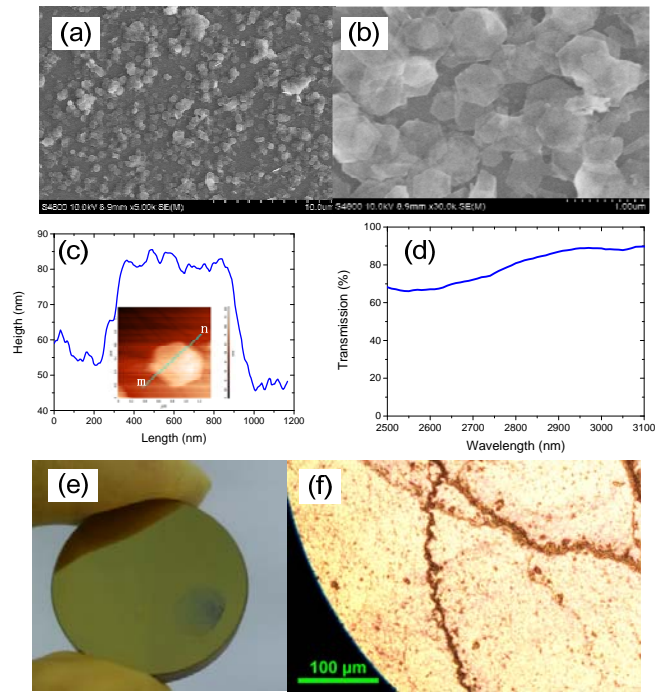

Fig. 1 Characteristics of the TI $Bi_2Te_3$ nano-sheets and the TI-SAM. (a) Low and (b) high magnification SEM images of the $Bi_2Te_3$ nano-sheets. (c) Height profile of a typical flake. (d) Broad transmission of $Bi_2Te_3$ nano-sheets. (e) Photo of the TI-SAM. (f) View of the TI

deposition area of the TI-SAM with an optical microscope. Inset of (c) shows the AFM image of the Bi$_2$Te$_3$ flake.

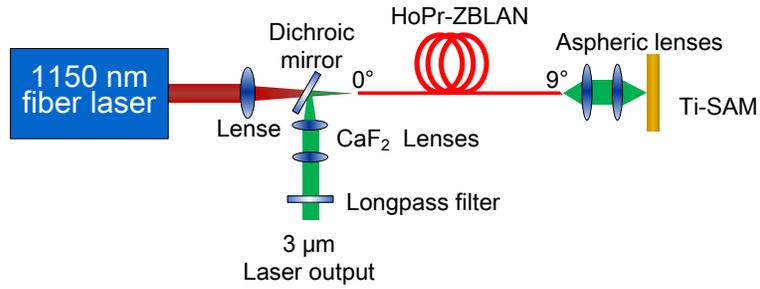

Fig. 2 Schematic of the mid-infrared mode-locked ZBLAN fiber laser with TI-SAM.

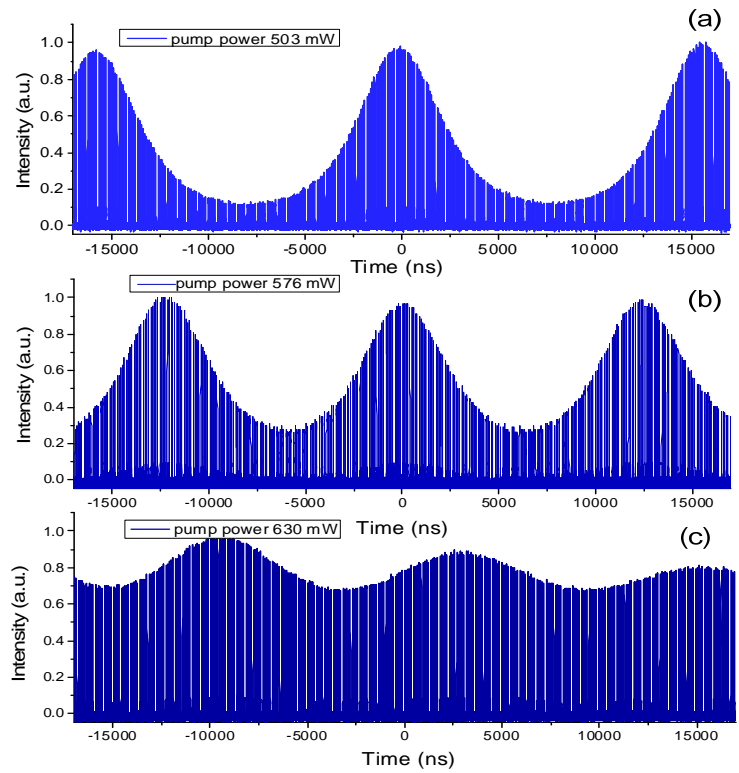

Fig. 3 QS-ML pulse trains under pump powers of (a) 503 mW, (b) 576 mW and (c) 630 mW.

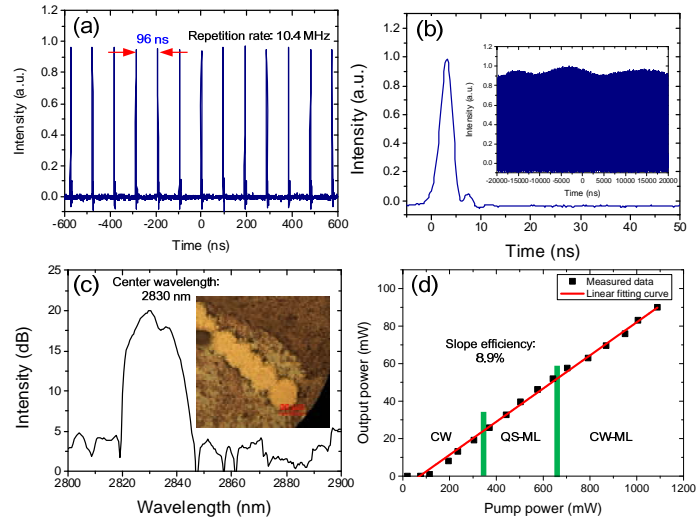

Fig. 4 Characteristics of the CW-ML pulses output. (a) CW-ML pulse train. (b) Single pulse shape. (c) Laser spectrum. (d) Evolution of the laser output power and operation states versus the pump power. Inset of (b) shows the CW-ML pulse trains over a large scale. Inset of (c) presents the damaged surface of the TI-SAM.